# Anomalous thermal transport and high thermoelectric performance of Cu-based vanadate CuVO$_3$

Xin Jin[1,2], Qiling Ou[1], Haoran Wei[3], Xianyong Ding[3], Fangyang Zhan[3], Rui Wang[3,4], Xiaolong Yang[3,4,*], Xuewei Lv[2,*], and Peng Yu[1,*]

[1]College of Physics and Electronic Engineering, Chongqing Normal University, Chongqing 401331, P. R. China.

[2]College of Materials Science and Engineering, Chongqing University, Chongqing 400044, P. R. China.

[3]College of Physics & Chongqing Key Laboratory for Strongly Coupled Physics, Chongqing University, Chongqing 400044, P. R. China.

[4]Center of Quantum materials and devices, Chongqing University, Chongqing 400044, P. R. China.

* Corresponding Author: yangxl@cqu.edu.cn, lvxuewei@163.com, pengyu@cqnu.edu.cn

# ABSTRACT

Thermoelectric (TE) conversion technology, capable of transforming heat into electricity, is critical for sustainable energy solutions. Many promising TE materials contain rare or toxic elements, so the development of cost-effective and eco-friendly high-performance TE materials is highly urgent. Herein, we explore the thermal transport and TE properties of transition metal vanadate $CuVO_3$ by using first-principles calculation. On the basis of unified theory of heat conduction, we uncover the hierarchical thermal transport feature in $CuVO_3$, where wave-like tunneling makes a significant contribution to the lattice thermal conductivity ($\kappa_l$) and result in the anomalously weak temperature dependence of $\kappa_l$. This is primarily attributable to the complex phononic band structure caused by the heterogeneity of Cu-O and V-O bonds. Simultaneously, we report a high power factor of 5.45 mW·K$^{-2}$·m$^{-1}$ realized in hole-doped $CuVO_3$, which arises from a high electrical conductivity and a large Seebeck coefficient enabled by the multiple valleys and large electronic density of states near the valence band edge. Impressively, the low $\kappa_l$ and the high power factor make $p$-typed $CuVO_3$ have $ZT$ of up to 1.39, with the excellent average $ZT$ above 1.0 from 300 to 600 K, which is superior to most reported Cu-based TE materials. Our findings suggest that $CuVO_3$ compound is promising candidate for energy conversion applications in innovative TE devices.

Amidst the global agenda of energy transition and climate change, the quest for materials that can efficiently convert energy forms is an urgent task. Thermoelectric (TE) materials, capable of directly converting waste heat into electrical energy, have emerged as an efficient solution for enhancing energy utilization without increasing the carbon footprint.[1-5]. The efficiency of TE materials in energy conversion is commonly evaluated using a dimensionless figure of merit, denoted as $ZT$[6]. This figure is defined by the formula $ZT = (S^2\sigma T)/(\kappa_e + \kappa_l)$, where $S$ represents the Seebeck coefficient, $\sigma$ is the electrical conductivity, $T$ is the absolute temperature, $\kappa_e$ is the electrical thermal conductivity, and $\kappa_l$ is the lattice thermal conductivity[7]. Achieving superior performance in TE materials typically demands the concurrent enhancement of the power factor ($PF = S^2\sigma$) and reduction of the total thermal conductivity ($\kappa = \kappa_e + \kappa_l$).

Over the past two decades, significant strides have been made in the development and innovation of high-efficiency TE materials. A variety of strategies have been employed to improve their TE performance, including the engineering of full-scale microstructure[8,9], manipulation of the electronic band structure[10], optimization of carrier concentration[11], and decoupling of electron and phonon transport[12,13]. Besides, achieving a high level of conversion efficiency still requires an exceptional average $ZT$ value across a varied operating temperature range. While TE materials like SnSe display impressive $ZT$ values at higher temperatures, above 750 K[14], others such as $Be_2Te_3$ excel in lower temperature ranges around 300 K[15,16]. However, the optimal performance of these materials at specific temperature ranges does not fully address the requirements for applications in environments experiencing wide temperature fluctuations. Therefore, it is imperative to develop high-performance TE materials that possess a broad operational temperature range to meet diverse application needs.

In recent years, the search for new materials with outstanding TE properties has garnered significant research interest[15-17]. Metal vanadates, in particular, have attracted attention due to their unique bond heterogeneity, which leads to low lattice thermal conductivities, as highlighted in several of our previous works[18,19]. In addition, the electronic structure of vanadates exhibits a large electronic density of states near the Fermi level and the characteristic of multiple valleys[19], which will be beneficial for the excellent electrical transport properties, such as a large Seebeck coefficient. These prominent characteristics suggest the potential of vanadates as promising TE candidates. Meanwhile, copper-based materials have also garnered substantial interest within the TE community, owing to copper's non-toxicity and earth-abundance, as well as the unique and unusual physical properties observed in some of these compounds[20,21]. Typical examples include $BiCuSeO$[22,23], $Cu_{12}Sb_4S_{13}$[24], and $Cu_2TiTe_3$[25], among others. In this context, developing copper-based vanadates with superior TE performance is a topic worth researching in the TE field. In this work, we study the lattice thermal transport and TE properties of Cu-based vanadate $CuVO_3$, by means of first-principles based Boltzmann transport equation (BTE) combined with the unified theory of lattice dynamics. Due to the complexity of crystal structure, $CuVO_3$ exhibits many flat phonon branches and soft modes, leading to hierarchical thermal transport behavior. The predicted $\kappa_L$ based on the unified theory is as low as 2.4 W/mK at room temperature, and shows weak dependence on temperature due to the significant contribution of diffusive transport. Moreover, we predict that $CuVO_3$ exhibits excellent TE performance with a peak *ZT* value of 1.39, which arises largely from the high power factor due to multiple valleys and large electronic density of states near the band edge. This work provides guidance for searching high-performance TE materials in vanadate compounds.

All density functional theory (DFT) calculations were implemented in the *Vienna Ab initio* Simulation Package (VASP)[26,27]. The interactions between ions and electrons were described by the Projector Augmented Wave (PAW) method[28], and the exchange-correlation effects were accounted for within the generalized gradient approximation (GGA) in the form of Perdew-Burke-Ernzerhof (PBE) functional[29,30]. To refine the description of electronic and phonon transport properties, the DFT+*U* approach was adopted[31,32], setting the effective Hubbard *U* parameters for V and Cu elements at 3 eV and 4 eV, respectively. The Brillouin zone sampling was determined using the Monkhorst-Pack scheme[33]. The cutoff energy for plane-wave basis was consistently set to 520 eV across all calculations. The energy and force convergence criteria were set to $10^{-8}$ eV and $10^{-4}$ eV/Å, respectively. Phonon dispersion calculations were performed on a 3×3×3 (270 atoms) supercell utilizing the PHONOPY software package[34]. Third-order force constants were computed within 3×3×3 supercell using the THIRDORDER code[35], wherein the interatomic interactions were considered up to the eighth nearest neighbors. The $\kappa_l$ was calculated via the dual-phonon model[36], by solving the phonon BTE with the modified ShengBTE package[35,37]. For achieving the convergence of thermal conductivity, the $q$-grid mesh was set to 15×15×15. After obtaining accurate electronic band structures, the BoltzTraP2 software package[38] was utilized to compute the electronic transport properties of $CuVO_3$.

**Fig. 1(a)** illustrates the crystallographic structure of $CuVO_3$. The compound crystallizes in the trigonal system, conforming to an ilmenite structure with the designated space group $R\bar{3}$. The lattice constants are determined to be: $a$ = 5.003 Å, $b$ = 5.003 Å, and $c$ = 14.317 Å. Within this framework, Cu and V atoms are situated at the 6$c$ Wyckoff positions, while O atoms are placed at the 18$f$ Wyckoff positions. The optimized lattice parameters exhibit commendable congruence with the experimental values reported in literature[39]. A closer inspection of the crystalline structure reveals that $CuVO_3$ is comprised of $VO_6$ and $CuO_6$ octahedra, both manifesting as distorted and irregular geometries. These polyhedra are intricately connected through the O atoms. Notably, the V-O bonds span lengths of 1.82 Å and 2.07 Å, whereas the Cu-O bonds extend to 2.28 Å. This bonding heterogeneity is responsible for the softening of phonon modes and low phonon group velocities and thus low $\kappa_l$ as discussed later.

Phonon dispersion curve reflects the intrinsic characteristic of atomic vibrations within crystals, dictating phonon thermal transport properties. **Fig. 1(b-c)** show the first Brillouin zone and phonon spectrum along the high symmetry directions as well as the projected phonon density of states (PDOS). The absence of imaginary phonon modes indicates the dynamic stability of $CuVO_3$. Similar to the phonon structures of other vanadate materials[18], the acoustic phonons of CuVO$_3$ exhibit soft modes near 2.5 THz, predominantly distributed along the Γ→C, P1→C, and C→L directions, implying the relatively low phonon group velocities. Moreover, there is significant overlap between acoustic branches and low-frequency optical branches, which could provide large three-phonon scattering channels for the heat-carrying acoustic phonons. From the PDOS, it is seen that the low-frequency acoustic phonons are primarily contributed by the heavier Cu atoms, and the hybridization between the density of states of V atoms and O atoms in the frequency range of 0–5 THz is markedly evident, whereas there is virtually no overlap with the Cu atoms. This indicates substantial differences in the vibrational modes between various chemical bonds within the system, making the acoustic and low-frequency optical phonons more susceptible to scattering, thereby significantly impeding their contribution to thermal transport. The mid-to-high frequency region is predominantly contributed by V and O atoms, owing to the disparity in atomic mass and the stronger V-O bonds. The analysis of the aforementioned phonon structures is of paramount significance for understanding the lattice thermal transport properties of CuVO$_3$.

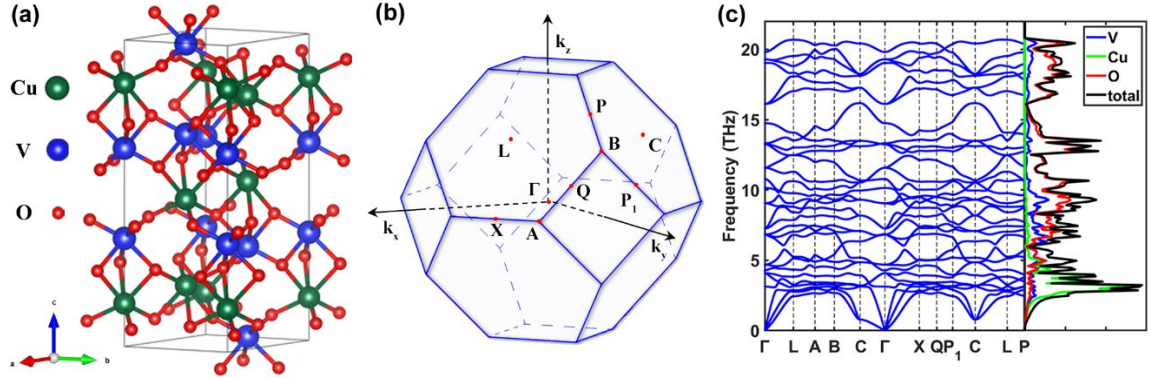

**Fig. 1** (a) Conventional crystal structure of CuVO$_3$, and the Cu, V and O atoms are shown as green, blue and red spheres, respectively. (b) The first Brillouin zone of CuVO$_3$. (c) Phonon dispersion curves along high-symmetry directions of the Brillouin zone and corresponding phonon density of states of CuVO$_3$.

Upon obtaining the phonon spectrum, the phonon transport properties of $CuVO_3$ are calculated using standard anharmonic lattice dynamics. As illustrated in **Fig. 2(a-c)**, a significant number of vibrational modes along the *a*, *b,* and *c* directions have mean free paths smaller than the Ioffe-Regel limit in space (defined by the minimum of interatomic spacing), indicating that large amounts of phonon modes cannot be considered as normally propagating phonons but as diffuson-like phonons. This also implies that conventional phonon BTE cannot fully explain the phonon transport behavior of $CuVO_3$. Recently, the unified heat conduction theories that include particle-like conduction and wave-like tunneling have been established and widely applied to explain the heat transport behavior of crystals and glasses[40-42]. Here, to account for the contribution of diffusive phonon transport, the $\kappa_l$ of $CuVO_3$ is calculated by using dual-phonon theory[36] and Wigner formulation[40], respectively. **Fig. 2(d-f)** depict the *T*-dependent $\kappa_l$ of $CuVO_3$ along different axes. It is observable that with increasing temperature, the dominance of propagative heat conduction progressively diminishes while diffusive transport channels become significant. Notably, the Wigner equation predicts the diffuson-like conductivity clearly different from that given by the dual-phonon model, especially at higher temperatures. The reason for the difference is twofold: (i) The latter considers diffusive phonon transport within the harmonic approximation while the former accounts also for anharmonicity; (ii) The two approaches have different criteria for defining diffuson-like phonons as described below. The first reason indicates that the Wigner formulation is more reliable in describing the diffusive transport, and hence we eventually adopt it to calculate the $\kappa_l$ of $CuVO_3$. At 300 K, the $\kappa_l$ predicted by the Wigner formulation along the *a*, *b*, and *c* axes are 1.9, 2.6, and 2.8 W/mK, respectively, with diffusive channels contributing 11.0%, 11.8%, and 11.7% accordingly. As the *T* rises to 600 K, contributions from diffuson-like phonons are notably enhanced, accounting for 0.61, 0.46, and 0.43 W/mK along the *a*, *b*, and *c* axes, respectively. Remarkably, the increasingly important role of diffusive transport channels with temperature results in an anomalously weak *T*-dependence of $\kappa_l$, particularly evident for the *a* axis. It should be noted that the room-temperature $\kappa_l$ values of $CuVO_3$ predicted by the Wigner formulation, are comparable to those of traditional TE materials such as $Bi_2Te_3$ (1.7 W/mK) and PbTe (2.3 W/mK)[43].

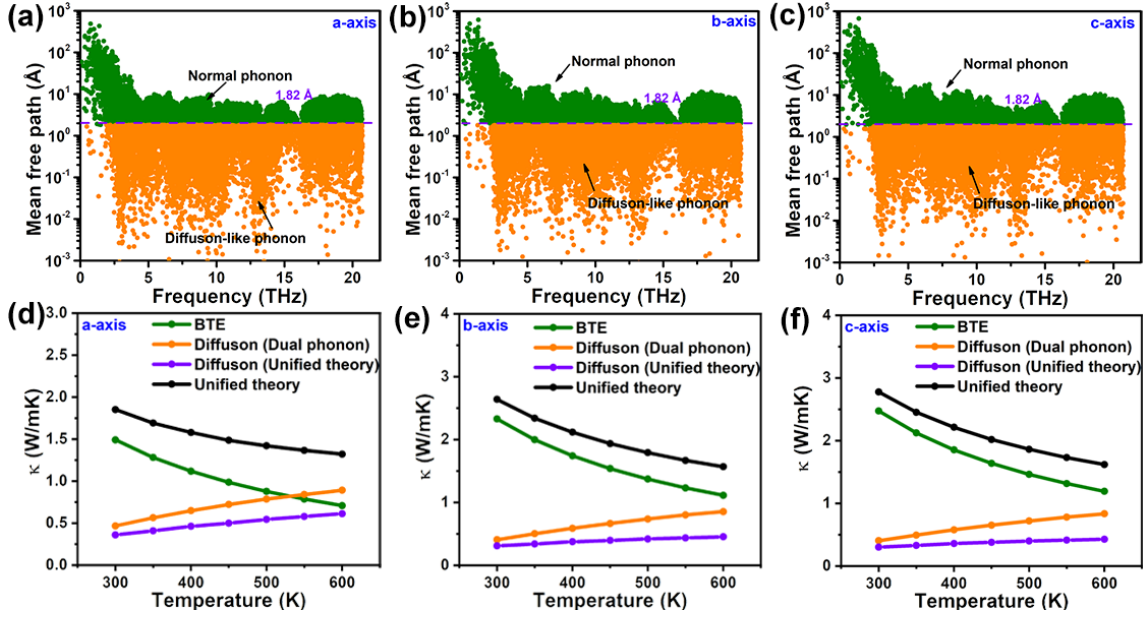

**Fig. 2** The calculated phonon mean free path at 300 K of $CuVO_3$ along (a) *a*, (b) *b* and (c) *c* axes. The horizontal dashed line marks the Ioffe-Regel limit in space. The temperature-dependent $\kappa_l$ along (d) *a*, (e) *b*, and (f) *c* axes, predicted from the different unified theories.

To seek the underlying mechanisms for the hierarchical thermal transport and low $\kappa_l$ of $CuVO_3$, we further investigate the modal phonon lifetime, group velocity, scattering phase space, and Grüneisen parameter. **Fig. 3(a)** displays the calculated phonon lifetimes, along with the Ioffe−Regel limit in time given by the Cahill−Watson−Pohl (CWP) model ($T_{\text{Ioffe−Regel}}$ is the inverse of the phonon frequency)[44] and the Wigner limit ($T_{\text{Wigner}}$ equals the inverse of the average interband spacing, defined as $T_{\text{Wigner}} = 1/\Delta\omega_{\text{ave}} = 3N_{at}/\omega_{\text{max}}$, where $N_{at}$ is the number of atoms in the primitive cell and $\omega_{\text{max}}$ is the maximum frequency)[45]. It is evident that a large number of vibrational modes have lifetimes longer than the Ioffe−Regel limit but shorter than the Wigner limit, indicating that most of the well-defined-in-time phonons defined within the dual-phonon theroy are actually wave-like diffusons according to the Wigner equation. It also suggests that a huge proportion of diffusons revealed in the hierarchical mean free paths should originate mainly from the relatively low phonon group velocity rather than the phonon lifetime. Indeed, as presented in **Fig. 3(b)**, the phonon group velocities above 2.5 THz are relatively low, which is closely associated with the bonding heterogeneity-induced soft modes and flat bands in the phonon spectrum. From the perspective of dual-phonon theory, we note that many reported crystalline materials exhibit hierarchical thermal transport behavior primarily owing to the large scattering rates caused by strong anharmonicity, such as $Tl_3VSe_4$[46] and $La_2Zr_2O_7$[36], whereas here the same behavior mainly arises from the relatively low group velocities. Although the complex phonon dispersion relation gives rise to the large three-phonon scattering phase space in $CuVO_3$ due to easily satisfied energy and momentum selection rules, as seen in **Fig. 3(c)**, which is comparable to that of $La_2Zr_2O_7$ in magnitude, the lattice anharmonicity revealed by the modal Grüneisen parameter (**Fig. 3(d)**) is not strong compared to $La_2Zr_2O_7$. Note that the average value of Grüneisen parameter in $CuVO_3$ is 1.49 at 300 K, far below the average value of 1.83 in $La_2Zr_2O_7$ and 7.2 in $SnSe$[14]. As a result, the intrinsic phonon-phonon scattering rates of most modes, inverse of phonon lifetimes, are consistently lower than their phonon frequencies which correspond to the Ioffe-Regel limit given by the CWP model.

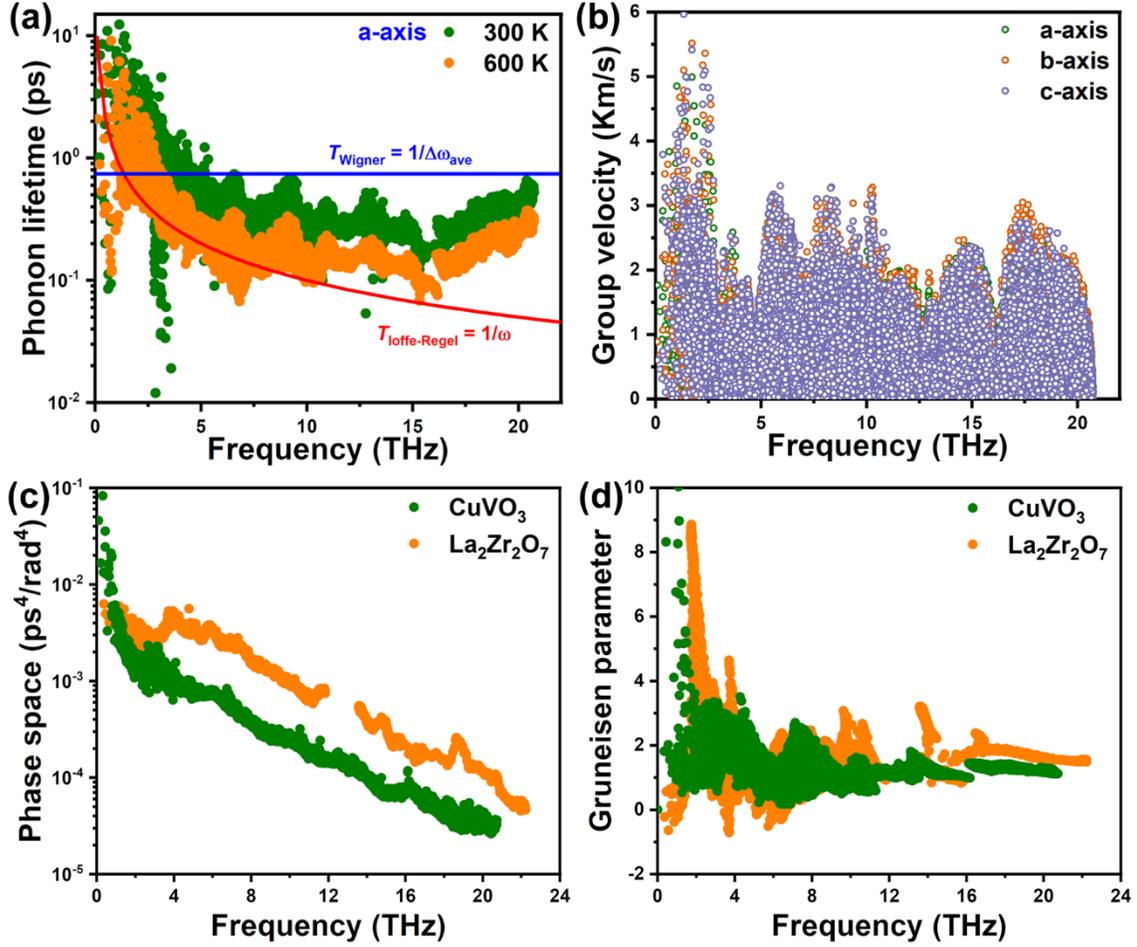

**Fig. 3** (a) The calculated phonon lifetimes at different temperatures. The solid blue and red lines represent the Wigner limit and Ioffe−Regel limit, respectively. (b) Phonon group velocities along the $a$, $b$, and $c$ axes. (c) Mode-resolved weighted three-phonon scattering phase space and (d) Grüneisen parameters of $CuVO_3$ and $La_2Zr_2O_7$.

The low $\kappa_l$ of $CuVO_3$ makes it a promising TE material, provided that its Seebeck coefficient $S$ and electrical conductivity $\sigma$ are sufficiently high. Given the high correlation between the electronic transport properties and electronic structure, it is imperative to calculate the electronic band structure and projected DOS of $CuVO_3$. Previous studies[19,47] have substantiated that the introduction of a Hubbard $U$ parameter in first-principles calculations can accurately characterize the band gap of vanadates. As depicted in **Fig. 4(a)**, our calculation shows that $CuVO_3$ is an indirect band gap semiconductor with a band gap of 0.63 eV, and possesses multiple electronic valence (conduction) bands with basically the same band maximum (minimum), which is beneficial for electronic transport properties. We also find from the electronic DOS that in the vicinity of the valence band maximum (VBM), the band structure is largely derived from the 3$d$ orbitals of Cu atoms, while near the conduction band minimum (CBM), the band structure is primarily dominated by the 3$d$ orbitals of V atoms. In contrast, the electronic DOS near the VBM is larger and steeper than that in the proximity of the CBM, indicating the larger DOS effective mass at the top of the valence band compared to the bottom of the conduction band. As a consequence, at the same carrier concentration, $p$-type doping tends to have a larger $S$ values, since the Seebeck coefficient is proportional to the DOS effective mass according to the Pisarenko relation[48].

We then investigate the electrical transport properties of CuVO$_3$ with the aid of semiclassical electron BTE. The calculated Seebeck coefficient, electrical conductivity, and power factor of $p$-type and $n$-type CuVO$_3$ at different doping concentrations are shown in **Fig. 4(b-c)** and **Fig. S1**. Note that as the transport properties along the three axes are basically the same, here the results along the $a$ axis are only provided. Notably, the peak value of the Seebeck coefficient for CuVO$_3$ exceeds 600 µV/K at 300 K, surpassing traditional Cu-based TE materials such as Cu$_2$S (250~300 µV/K)[49], BiCuSeO (260 µV/K)[50], and SnS$_{0.91}$Se$_{0.09}$ (~430 µV/K)[51]. The high Seebeck coefficient can be attributed to a large DOS effective mass near the Fermi level caused by multiple valleys. Also, as expected, the Seebeck coefficient for $n$-type doping is slightly lower than that for $p$-type doping as a result of the larger DOS effective mass near the VBM. It is noteworthy that the electrical conductivity for both $p$-type or $n$-type CuVO$_3$ is up to 10$^5$-10$^6$ Ω$^{-1}$·m$^{-1}$, which would endow it with a relatively high power factor. Given that the electronic thermal conductivity and electrical conductivity follow the Wiedemann-Franz-Lorenz's law[52], $\kappa_e = L\sigma T$, where $L$ denotes the Lorenz number, the electronic thermal conductivity exhibits variation trend with temperature and carrier concentration similar to $\sigma$. As illustrated in **Fig. S2**, the electronic thermal conductivity becomes non-negligible only when the carrier concentration exceeds $1\times10^{21}$ cm$^{-3}$. Since the $S$ and the $\sigma$ exhibit opposite trends with respect to the carier density, there must exist an optimal carrier concentration that maximizes the power factor ($PF = S^2\sigma$). As seen in **Fig. 4(d)**, the calculated peak value of $PF$ for $p$-type CuVO$_3$ reaches 5.45 mW·K$^{-2}$·m$^{-1}$ at 300 K when the carrier concentration reaches $1.2\times10^{21}$ cm$^{-3}$, exceeding Pd$_2$Se$_3$ (1.21~1.61 mW·K$^{-2}$m$^{-1}$)[53], PbTe (~2.0 mW·K$^{-2}$m$^{-1}$)[54], and comparable to Bi$_2$Te$_3$ (~4.5 mW·K$^{-2}$m$^{-1}$)[55]. Hence, it is anticipated that CuVO$_3$ could achieve an excellent level of $ZT$ value.

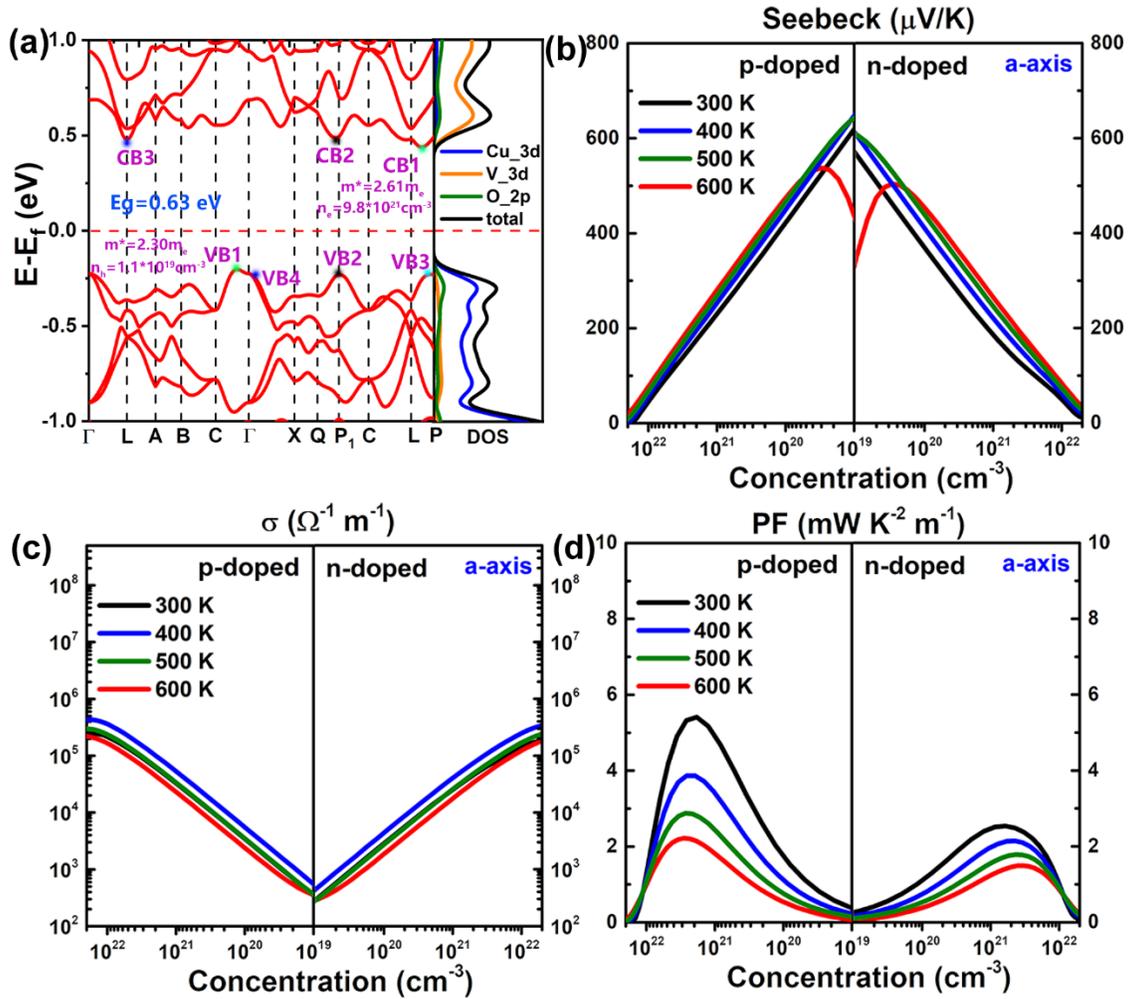

**Fig. 4** (a) Electronic band structure along high-symmetry directions of the Brillouin zone and corresponding density of states of CuVO$_3$. (b) Seebeck coefficient, (c) electrical conductivity, and (d) power factor (*PF*) of CuVO$_3$ system along *a*-axis as a function of carrier concentration at 300, 400, 500 and 600 K.

Upon acquiring all the electrical and thermal transport properties, *ZT* of CuVO$_3$ material is calculated to assess its TE performance, as shown in **Fig. 5(a)**. The results indicate that due to a high power factor and low $\kappa_l$, CuVO$_3$ exhibits excellent TE performance along the *a*, *b*, and *c* axes. Particularly in the a-direction, the highest *ZT* value is observed due to its lower $\kappa_l$. At room temperature, the *ZT* value for *p*-type CuVO$_3$ reaches 0.83 with a carrier concentration of $1.2 \times 10^{21}$ cm$^{-3}$. At elevated temperatures, the *ZT* peak value for *p*-type CuVO$_3$ increases to 1.39 at 600 K, surpassing Cu-based TE materials such as Cu$_5$Sn$_2$S$_7$ (0.45)[56], Cu$_{3.21}$Ti$_{1.16}$Nb$_{2.63}$O$_{12}$ (0.72)[57], Cu$_3$SbS$_4$ (0.77)[58], and is comparable to BiCuSeO (1.05)[59]. Notably, an excellent TE material requires high *ZT* values across the entire temperature range to achieve efficient TE conversion efficiency. However, traditional TE materials like SnSe exhibit high *ZT* values above 750 K[14], while Be$_2$Ti$_3$ demonstrates high *ZT* values at lower temperature regimes (~300 K)[15]. The superior performance of these materials at specific temperatures does not meet the needs for broad applications in environments with significant temperature fluctuations. To provide a comprehensive performance assessment, the average *ZT* value from 300 to 600 K is also calculated. As depicted in **Fig. 5(b)**, the TE performance of CuVO$_3$ varies relatively little with temperature, consistently maintaining a high level, implying robust TE conversion efficiency across a wide temperature range, thus holding significant application potential in environments with large temperature variations. More importantly, the peak *ZT* possesses a higher value among the Cu-based TE materials, as shown in **Fig. 5(c)**[50,56-64]. These results not only provide valuable predictive guidance for further experimental investigation into the TE properties of CuVO$_3$ but also enhance the understanding of the performance of TE materials at varying temperatures. Furthermore, these findings may inspire research into other materials with similar structures or properties, laying a foundation for developing more high-performance TE materials.

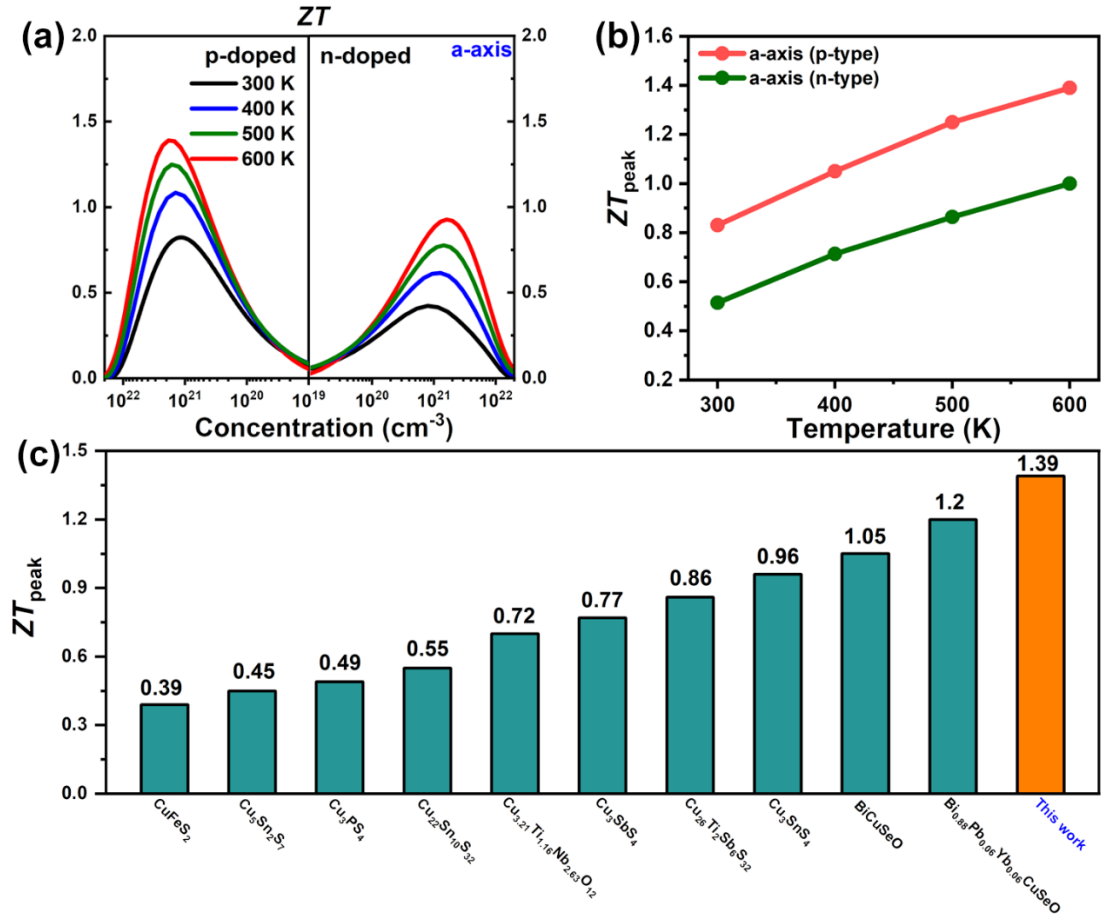

**Fig. 5** (a) *ZT* and (b) *ZT* peak value of *n*-type and *p*-type doping along *a* axis in CuVO$_3$ material at different temperature. (c) Comparison of this work to Cu-based eco-friendly TE materials in terms of peak *ZT*[50,56-64].

In summary, we have conducted a comprehensive study of phonon transport and TE properties of CuVO$_3$. As a result of complex phonon dispersion structure induced by the bonding heterogeneity of Cu-O and V-O bonds, CuVO$_3$ emerges hierarchical thermal transport behavior, where diffusive channels play a important role in determining $\kappa_l$ and lead to the abnormally weak temperature dependence of $\kappa_l$. Accounting for the contributions of normal and diffuson-like phonons, the predicted $\kappa_l$ at room temperature is 2.4 W/mK. Meanwhile, the presence of multiple valleys and large electronic DOS near the band edge endows CuVO$_3$ with high electronic conductivity and large Seebeck coefficient, leading to the high power factor of 5.45 mW·K$^{-2}$·m$^{-1}$ in $p$-type case. At optimal carrier concentrations, $p$-type CuVO$_3$ shows excellent TE performance with a peak $ZT$ value of 1.39 and the average $ZT$ of 1.0 from 300 to 600 K. This work highlights the potential of CuVO$_3$ in advanced TE applications and offers an indicator for the discovery of high-performance TE materials in vanadate compounds.

See the supplementary material for more details about the unified theory, Seebeck coefficient, electrical conductivity, power factor, and electronic thermal conductivity along different axes of CuVO$_3$.

This work is supported by the Natural Science Foundation of China (NSFC) (Grant Nos. 12374038, 12147102, 52071043, and 52371148), the National Key Research and Development Program of China (No. 2022YFC3901001-1), the Foundation Research Fund for the NSFC (U1902217), the Foundation of Chongqing Normal University (23XLB015, 21XLB046), and the Science and Technology Research Program of Chongqing Municipal Education Commission (No. KJQN-202200510). We gratefully acknowledge HZWTECH for providing computation facilities.

# AUTHOR DECLARATIONS

## Conflict of Interest

The authors have no conflicts to disclose.

## Author Contributions

**Xin Jin:** Investigation, Writing, original draft. **Qiling Ou**: Formal analysis, Methodology. **Haoran Wei:** Formal analysis, Validation. **Xianyong Ding:** Formal analysis, Methodology. **Fangyang Zhan:** Formal analysis, Methodology. **Rui Wang:** Methodology. **Xiaolong Yang:** Conceptualization, Formal analysis, Methodology, Writing, Funding acquisition. **Xuewei Lv:** Conceptualization, Funding acquisition. **Peng Yu:** Conceptualization, Funding acquisition.

# DATA AVAILABILITY

The data that support the findings of this study are available from the corresponding author upon reasonable request.